# Galvanomagnetic effects in graphene

I. I. Boiko[1]

*Institute of Semiconductor Physics, NAS of Ukraine, 45, pr. Nauky, 03028, Kiev, Ukraine*

(Dated: November 13, 2010)

Magnetoresistivity and Hole-effect were theoretically investigated for neutral and gated graphene. It is shown that in neutral graphene Hall-effect is totally absent. In gated, exactly monopolar graphene effect of magnetoresistivity vanishes; here Hall-constant does not involve any relaxation characteristic in contrast to result obtained for popular method of τ-approximation.

## Introduction

Parallel to search of conductivity the investigation of magnetotransport gives a lot of useful information about properties of studied materials. Various galvanomagnetic effects, especially Hall-effect and magnetoresistivity, are natural elements of consistent way for qualified search of kinetic coefficients in crystals at different temperatures, densities of carriers, etc. At present one of the most interesting subject of scientific research is graphene (see Refs. [1] – [4])

## 1. Quantum kinetic equation

Consider here 2D-system: an uniform graphene crystal in constant uniform electrical and magnetic fields $\vec{E}$ and $\vec{H}$. For this case the stationary quantum kinetic equation for the distribution function $f_{\vec{k}_\perp}^{(a)}$ of band carriers from *a*-group, moving in the plane $z=0$, can be presented in the form (see Ref. [5])

$$\frac{e_a}{\hbar}\vec{E}\frac{\partial f_{\vec{k}_\perp}^{(a)}}{\partial \vec{k}_\perp} + \frac{e_a}{\hbar}\left\{\frac{1}{c}\left[\left(\vec{H}\times\frac{\partial}{\partial \vec{k}_\perp}\right),\vec{v}_\perp^{(a)}(\vec{k}_\perp)\right]_+ f_{\vec{k}_\perp}^{(a)}\right\} = \text{St} f_{\vec{k}_\perp}^{(a)}. \qquad (a = e \text{ or } h). \qquad (1.1)$$

Here $\text{St} f_{\vec{k}_\perp}^{(a)}$ is collision integral. In this paper we consider classical magnetic field (for massless fermions it corresponds to such condition: $H << c(k_B T)^2/e\hbar v_F^2$). Farther we assume the following orientation of fields:

$$\vec{E} = \vec{E}_\perp = (E_x, E_y, 0) \; ; \; \vec{H} = (0,0,H_z). \qquad (1.2)$$

Dispersion laws and microscopic velocities of electrons and holes in neutral graphene are equal:

$$\varepsilon^{(e,h)}(\vec{k}_\perp) = \hbar v_F k_\perp \quad ; \quad \vec{v}^{(e,h)}(\vec{k}_\perp) = \frac{1}{\hbar}\frac{\partial \varepsilon^{(e,h)}(\vec{k}_\perp)}{\partial \vec{k}_\perp} = v_F \frac{\partial k_\perp}{\partial \vec{k}_\perp} = v_F \frac{\vec{k}_\perp}{k_\perp} . \qquad (1.3)$$

---

[1] Electron address: igorboiko@yandex.ru

Applying to both sides of Eq. (1) the operation

$$(1/\pi^2)\int \vec{k}_\perp d^2\vec{k}_\perp,$$

one obtains a set of exact balance equations for dynamic and statistic forces:

$$e_a[\vec{E}_\perp + (1/c)(\vec{H}\times\vec{u}_\perp^{(a)})] + \vec{F}_v^{(a)} = 0.  \qquad (1.4)$$

Here the vector

$$\vec{u}_\perp^{(a)} = \frac{\int \vec{v}_\perp^{(a)}(\vec{k}_\perp) f_{\vec{k}\perp}^{(a)} d^2\vec{k}_\perp}{\int f_{\vec{k}\perp}^{(a)} d^2\vec{k}_\perp} \qquad (1.5)$$

is drift velocity for *a*-group (see Refs. [5] and [6]). The vector $\vec{F}_v^{(a)}$ is a friction force, provoked by interaction of forced band carriers with microscopic scattering fields. Note, that for equilibrium state of carriers the drift velocities $\vec{u}^{(e,h)} = 0$, and collision integrals $\text{St} f_{\vec{k}\perp}^{(e,h)} = 0$, also. So long as collision integrals do not contain dynamic forces by downright way, there is nothing for linear theory to adopt the friction forces at equilibrium scattering system

$$\vec{F}_v^{(a)} = \frac{\hbar}{\pi^2 n_a}\int \vec{k}_\perp [\text{St} f_{\vec{k}\perp}^{(a)}] d^2\vec{k}_\perp$$

as directly proportional to drift velocity $\vec{u}_\perp^{(a)}$. For the similar reason we assume that macroscopic friction force, provoked by electron-hole scattering, is proportional to the difference $\vec{u}_\perp^{(e)} - \vec{u}_\perp^{(h)}$.

## 2. Galvanomagnetic effects in neutral graphene

In neutral graphene Fermi-level $\varepsilon_F$ lays exactly in the point $\varepsilon = 0$; so carriers densities $n^{(e)}$ and $n^{(h)}$ are equal:

$$n^{(e,h)} = \frac{2}{\pi}\left(\frac{k_B T}{v_F \hbar}\right)^2 \int_0^\infty \kappa[1+\exp(\kappa)]^{-1} d\kappa \approx 0.524 \left(\frac{k_B T}{v_F \hbar}\right)^2.  \qquad (2.1)$$

It was noted above, that total friction forces can be presented in the forms (see Ref. [7], also)

$$\vec{F}_v^{(e)} = -b\vec{u}_\perp^{(e)} - \varsigma(\vec{u}_\perp^{(e)} - \vec{u}_\perp^{(h)}) \;;\quad \vec{F}_v^{(h)} = -b\vec{u}_\perp^{(h)} - \varsigma(\vec{u}_\perp^{(h)} - \vec{u}_\perp^{(e)}).  \qquad (2.2)$$

Here kinetic coefficient $b$ is responsible for interaction of band carriers with external scattering system (impurities, phonons). Other kinetic coefficient $\varsigma$ is responsible for interaction between electrons and holes, which have different drifting velocities; this interaction in phenomenological form can be considered as mutual drag.

The definite expressions for $b$ and $\varsigma$ were obtained with the help of model distribution functions having the form of "shifted" Fermi-distribution (see Refs. [5] and [6]). For neutral graphene (see *e. g.* Ref. [7])

$$\varsigma \approx 1.386 e^3 (k_B T)^2 / \hbar^3 v_F^4 \varepsilon_L^2 \qquad (2.3)$$

and

$$b = 2\pi^2 e^3 n_{CI}^{(2)} / \varepsilon_L^2 \hbar\ v_F^2, \qquad (2.4)$$

if dominate external scattering system is presented by charged impurities disposed with density $n_{CI}^{(2)}$ in graphene plane.

From Eqs. (1.4) and (2.2) one obtains the following system (here and later $e > 0$):

$$-[\vec{E}_\perp + (1/c)(\vec{H} \times \vec{u}_\perp^{(e)})] - b\vec{u}_\perp^{(e)} - \varsigma(\vec{u}_\perp^{(e)} - \vec{u}_\perp^{(h)}) = 0\ ; \qquad (2.3)$$

$$[\vec{E}_\perp + (1/c)(\vec{H} \times \vec{u}_\perp^{(h)})] - b\vec{u}_\perp^{(h)} - \varsigma(\vec{u}_\perp^{(h)} - \vec{u}_\perp^{(e)}) = 0\ . \qquad (2.4)$$

Solution of this system:

$$\vec{u}_\perp^{(h)} = \frac{1}{(H_z^2/c^2 b + b + 2\varsigma)}[(1/bc)(\vec{H} \times \vec{E}_\perp) + \vec{E}_\perp]\ ;$$

$$\vec{u}_\perp^{(e)} = \frac{1}{(H_z^2/c^2 b + b + 2\varsigma)}[(1/bc)(\vec{H} \times \vec{E}_\perp) - \vec{E}_\perp]. \qquad (2.5)$$

Then density of current is

$$\vec{j}_\perp = \vec{j}_\perp^{(h)} + \vec{j}_\perp^{(e)} = en^{(e)}(\vec{u}_\perp^{(h)} - \vec{u}_\perp^{(e)}) = \frac{2en^{(e)}}{(H_z^2/c^2 b + b + 2\varsigma)}\vec{E}_\perp\ . \qquad (2.6)$$

It follows from Eq. (2.6), that in neutral graphene $\vec{j}_\perp \parallel \vec{E}_\perp$; that is, Hall-field, normal to the total current, does not arrives. What concerns transverse magnetoresistivity, it exists here:

$$\sigma(H_z) = \frac{2en^{(e)}}{H_z^2/c^2 b + b + 2\varsigma}\ ; \qquad \sigma(0) = \frac{2en^{(e)}}{b + 2\varsigma}\ . \qquad (2.7)$$

These two formulae show a possibility, to find two kinetic coefficients $b$ and $\varsigma$ with the help of not complicated experimental measurements. Note, that for high magnetic fields the dependence of conductivity on drag coefficient $\varsigma$ disappears.

### 3. Galvanomagnetic effects in gated graphene

Consider here graphene with monopolar conductivity created by applying of sufficiently high voltage $V_e$ to a gate ($eV_e \gg k_B T$). Let it will be *n*-graphene. Then the density of electrons is

$$n_e = \frac{1}{\pi^2}\int d^2\vec{k}_\perp \left[1+\exp\left(\frac{\varepsilon^{(a)}(k_\perp)-eV_e}{k_B T}\right)\right]^{-1} = \frac{2}{\pi}\left(\frac{k_B T}{v_F \hbar}\right)^2 \int_0^\infty \kappa[1+\exp(\kappa-\kappa_e)]^{-1}d\kappa. \quad (3.1)$$

Here $k_e = eV_e/k_B T > 0$. Number of holes is exponentially small, and we neglect contribution of holes in transport phenomena. In this case balance equation for electrons has the form (see Eq. (2.3))

$$-[\vec{E}_\perp + (1/c)(\vec{H}\times\vec{u}^{(e)})] - b_e \vec{u}^{(e)}_\perp = 0 \ . \quad (3.2)$$

The density of current is

$$\vec{j}_\perp = -en_e \vec{u}^{(e)}_\perp \ . \quad (3.3)$$

Solving the vector equation (3.2) one obtains:

$$u_x^{(e)} = -\frac{bE_x + (H_z/c)E_y}{b^2+(H_z/c)^2} \ ; \quad u_y^{(e)} = \frac{(H_z/c)E_x - bE_y}{b^2+(H_z/c)^2} \ . \quad (3.4)$$

Let the current is directed along $x$-axis: $\vec{j}_\perp = (j_x, 0)$. Determine Hall-constant $R_H$ and magnetic conductivity $\sigma_H$ by the relations

$$E_y = R_H H_z j_x \ ; \quad j_x = \sigma_H E_x \ . \quad (3.5)$$

It follows from Eqs. (3.3) – (3.5), that for arbitrary strength of magnetic field

$$\sigma_H = \frac{en^{(e)}}{b} \ ; \quad R_H = \frac{1}{en^{(e)}c} \ . \quad (3.6)$$

One can see from here that effect of transverse magnetoconductivity of exactly monopolar graphene does not manifest (the conductivity is not dependent on magnetic field). Hall-constant does not contain any values responsible for rate of momentum relaxation.

Note, that formulae (3.7) are distinct on well-known formulae, obtained with the help of standard $\tau$-approximation. In particular, deduced here Hall-constant $R_H$ does not involve in its structure so-called "Hall-factor", which depends on some specific time of relaxation. Appearance of latter factor in literature is totally connected with the principal imperfection of method of $\tau$-approximation (see, for instance, Ref [8]).

Our consideration was grounded on small number of assumptions, and they were not too hard. So obtained results have a sufficiently high level of reliability.

---